\documentclass[conference,letter,IEEEptsizeten,10pt]{IEEEtran}
\IEEEoverridecommandlockouts
\usepackage{cite}

\ifCLASSINFOpdf
\else
\fi
\usepackage{url,amsmath,graphicx,psfrag,color}

\begin{document}
%
\title{Interference Alignment (IA) and Coordinated Multi-Point (CoMP) overheads and RF impairments: testbed results 
\thanks{This work was done within the framework of the 
Swedish SSF sponsored RAMCOORAN project and the EU
project RAMCOORAN. 
The HiATUS project acknowledges the financial support of the
Future and Emerging Technologies program within FP7 for Research
of the European Commission (FET Open grant number 265578).}}

\author{\IEEEauthorblockN{Per Zetterberg}
\IEEEauthorblockA{Access Linnaeus Center\\
KTH Royal Institute of Technology, Osquldas v{\"{a}}g 10,\\
SE-100 44 Stockholm, Sweden, \\
Email: perz@ee.kth.se}}


%


\maketitle

\begin{abstract}

In this work we investigate the network MIMO techniques of interference
alignment (IA) and joint transmission coordinated
multipoint (CoMP) in an indoor very small cell environment.
Our focus is on the overheads in a system with quantized channel state
feedback from the receiver to the transmitter (based on the 802.11ac
standard) and on the impact
of non-ideal hardware.  
The indoor office scenario should be the most favorable
case in terms of the required feedback rates due to 
the large coherence bandwidth and coherence time of the channel.
The evaluations are done using a real-world wireless testbed
with three BSs and three MSs all having two antennas.
The signal to noise ratio in the measurements is very high, 35-60dB,
due to the short transmission range. Under such conditions
radio hardware impairments becomes a major
limitation on the performance. We quantify the impact of these
impairments. 
For a 23ms update interval the overhead is 2.5\% and
IA and CoMP improves the sum throughput 27\% and 47\% in average
(over the reference schemes e.g. TDMA MIMO), under stationary
conditions. When two people are walking in the measurement area
the throughput improvements drops to 16\% and 45\%, respectively.
\end{abstract}


%
\IEEEpeerreviewmaketitle

\section{Introduction}

Network MIMO techniques has been the focus
of much research interest, see e.g. \cite{GES:10}.  
An overview of some 3GPP work and results in the area
can be found in \cite{Lee12}. The feedback schemes standardized in 3GPP
are focused on reporting channel quality information (CQI)
and precoder matrix index (PMI) of a few transmission and interference
hypotheses rather than providing full channel state information
(channel matrices) to the transmitters, \cite{Lee12}.
 In much of the more academic work in the area, 
full channel state information CSI is a common assumption.
Such full information incurs a high overhead that may
be hard to justify in outdoor scenarios with significant mobility.
In this work we focus on an indoor very dense scenario where
full CSI may still be a worthwhile option to pursue.
The latest WiFi standard 802.11ac, aimed at such environments
incorporates a feedback scheme
which provides the transmitter with almost full channel state information
for the MU-MIMO scenario (single access point serving multiple users).
Herein we use the feedback compression scheme of that standard 
in a network MIMO context in the form of coordinated multipoint (CoMP) and 
interference alignment (IA). Although we use a feedback scheme from the WiFi
domain, a similar scheme could be introduced also in cellular systems aimed
at the same environment.

A breakthrough in information theory was achieved
with ``interference alignment'' which was presented in the paper \cite{CAD:08}.
According to the theory, $K/2$, interference free modulation streams
can be created in a system with $K$ links ($K$ transmitters and
$K$ receivers), even with only a single antenna at each node. 
The catch is, to achieve
this capacity; linear precoding over multiple channel realizations
(channel extensions)
is needed besides global channel knowledge. The most practical
implementation of channel extensions is probably
to utilize the subcarriers of an OFDM modulation thereby creating
diagonal MIMO channels. However the required number of subcarriers 
grows extremely fast with $K$. 
Results under more realistic assumptions
have so far been modest \cite{BRA:13}. 

Interference alignment (IA) can also be applied without
channel extensions using multiple antennas (i.e. on MIMO channels)
in a way 
that is in fact a version of coordinated beamforming,
see e.g. \cite{YET:10,TRE:09}.
The interference alignment theory in \cite{CAD:08},
provides us with 
a closed solution to the problem of creating
$K M/2$ interference free modulation streams in a system with $K=3$ links
where each node has $M$ antennas. 

In this paper we study the scenario with three BSs and three MSs
with two antennas each. We use the MIMO version of interference
alignment as described above.
In addition we also investigate a form of coordinated multipoint
transmission where all six transmitter antennas in the system
are used collectively as a single coherent antenna array to optimize 
the signal to interference ratio of all users.

This paper includes not only simulation
results but also measurements. Moreover, we do not just perform 
simulations based on channel measurements but present over-the-air
actual transmissions over the channel.
By doing so we establish
a ``ground truth'' of what is achievable with IA and CoMP.
Note that the true wireless channel from the base-band of
the transmitter to the base-band of the receiver
does not only involve the radio
propagation channel but also the impact of the analog hardware of the 
transmitter
and receiver. This effect includes e.g. the impact of amplifier non-linearities
and phase-noise, \cite{SCH:08}.
In this paper we use transmitters and receivers with error
vector magnitudes (EVM) of 1-4\%, see Section \ref{impair},
which is reasonable model of consumer-grade wireless equipment.

Previous work on experimentation within IA and CoMP includes the papers
\cite{AYA:10,GAR:11,GAR:11b,ZET:12a,ZET:14a,JUN:10,DON:11,HOL:11}.
A short summary of some of the main findings in these papers
are given in Section I of \cite{ZET:14a}.
In this paper, as in \cite{ZET:14a}, we
use the feedback compression scheme defined for MU-MIMO in 
the IEEE802.11ac standard but applied to IA and COMP,
in an indoor office environment. 
{\em This paper has the following novel aspects}, 
1) the impact of people walking
in the measurement environment (time varying channels) is studied 
experimentally, 2) the overhead of perfoming IEEE802.11ac feedback
is studied in IA and CoMP scenarios, 3) we factor out the impact of
hardware impairments (the methodology as well as the results)
, 4) we use an impairment model
based on look-up tables of error vector magnitude (EVM) versus
RMS signal strength and investigate its applicability,

The feedback update rate is 23ms and we investigate
the performance at the end of this update period. By using a combination
of measurements and simulations we also analyze a 3ms update interval,
in Section \ref{ima}. 
For our system the overhead, using the IEEE802.11ac standard, is 2.5\%;
see Section \ref{feedback}.
IA and CoMP improves the sum throughput 27\% and 47\% in average
(over the reference schemes e.g. TDMA MIMO), under stationary
conditions. When two people are walking in the measurement area
the throughput improvements drops to 16\% and 42\%, respectively.
Unfortunately, due to 
implementation difficulties related to computational loads
and system latencies in our PC based implementation, 
we are not able to reduce the update interval.
However, by using a combination of simulation and measurements we 
estimate that IA can improve its sum throughput from 11.0bits/sec/subcarrier
to 11.6bits/sec/subcarrier by using a 3ms update interval,
see Section \ref{ima}.

The {\em general conclusions} from the present paper is 
that IA and CoMP can provide an improvement over the reference schemes
even when considering a realistic feedback scheme and 
channel variations albeit in one of the most favorouble
scenarios. At the same time, the paper also shows 
that the performance is lower than what would be predicted
by simulations based on channel measurements in the same
environment. This highlights the necesity of considering hardware
impairments in system analysis. We also use a commonly employed
hardware impairment model. In order to get a good match between
simulation and measurements, we have made look-up tables of
EVM versus signal strength. Despite this effort, there is still
a discrepancy between measurement and simulation results. 
This result motivates further research into the issue and 
the use of over-the-air performance evaluations to obtain
truly realistic results. 
The results also show that CoMP is less sensitive than IA to channel variations and
hardware impairments, where both schemes use the same number of spatial streams.

The paper is organized as follows. In Section \ref{prel} we give
an overview of the testbed.
Section \ref{impl} describes the signal processing including the
MS to BS feedback mechanisms. The hardware impairment model
is described in Section \ref{impair}. 
The measurement results are provided in Section \ref{res}.
Finally, our conclusions are summarized in Section \ref{conc}.

\section{Hardware Overview}
\label{prel}

\begin{figure}
\centerline{
   \includegraphics[width=0.35\textwidth, height=0.223\textheight ]%
     {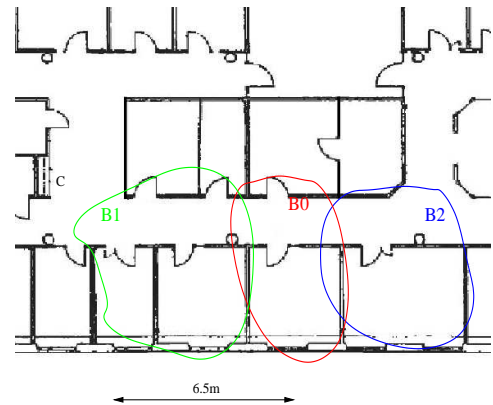}}
\caption{Map over the measurement area}
\label{C}
\end{figure}

\begin{figure}
\centerline{
   \includegraphics[width=0.3\textwidth, height=0.25\textheight ]%
     {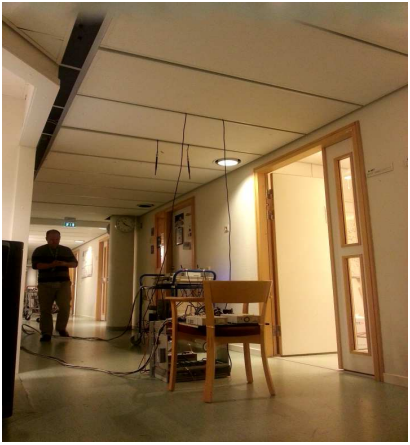}}
\caption{Picture of measurement setup}
\label{P}
\end{figure}

We consider a scenario with three links.
The propagation environment is indoor office. A map
of the measurement area is shown in Figure \ref{C}.
The positions of the three base-stations are marked
``B0'', ``B1'' and ``B2''. The red, green and blue area
shows the coverage area of each BS (i.e. the corresponding
MS was moved around in this area during the measurements,
each one of the three BSs easily covers all the areas
in terms of SNR). A picture taken from the location marked
``C'' in the map is shown in Figure \ref{P}. In the picture, the
base-station antennas of BS A are visible as well as
some of the testbed hardware.

The BSs consists of two USRP N210 with
XCVR2450 daughterboards (see \url{www.ettus.com}).
They are connected to 
mini-circuits (see \url{www.minicircuits.com})
ZRL-2400LN amplifiers which in turn are connected
to two vertically polarized antennas which are hanging down
from the ceiling. The base-band signals are generated by
a PC which are connected to the USRPs using ten meter
Gigabit Ethernet cables. 

The MSs also uses two USRP N210. Here we are however
using custom made RF front-ends; see \cite{ZET:03a}.
The base-band
signal is then sampled by the USRP N210s. The system operates at 2490MHz.
The receivers are also connected to a PC using ten meter Gigabit 
Ethernet cables. 

The processing of all three BSs runs on the same PC as
separate threads. Likewise all the three MSs runs
on another PC, as separate threads. The MS to BS feedback
is done over wired Ethernet, but compressed according to
Section \ref{feedback}. All the software is available as open source
at \url{sourceforge.net} under the project names {\tt fourmulti}
and {\tt iacomp}. 

The thermal noise standard deviation is roughly the same, 
$\sigma^2_{\text{nominal}}$, in all
receivers. This value is known by all nodes.

\section{Implementation}
\label{impl}

\subsection{Air Interface and Signal Processing Overview}

We use an OFDM modulation scheme with a subcarrier spacing of
312.5kHz, coinciding with that of 802.11a/n/ac.
The cyclic prefix is 0.4$\mu$s and thus the symbol time 3.6$\mu$s.
Due to limited sampling speeds of the USRP only 38 subcarriers
are used instead of the (at least) 58 used by 802.11n/ac. 
Two pilot subcarriers are inserted in each symbol at +7 and -7th
subcarriers from the center frequency,
and used for common phase error correction.

Ten different coding and modulation schemes (MCSs) have been implemented ranging
from one to six bit/symbol/subcarrier using QPSK to 256QAM constellations
and LDPC codes of rate 1/2 to 3/4. 

In each run of the system, two frames are transmitted. 
The first frame contains only six known training symbols
transmitted sequentially from the six BS antennas in the system.
This frame takes the function of the null data packet (NDP)
in MU-MIMO feedback of the IEEE802.11ac standard, see \cite{POR:13}.

The MSs estimate the $2{\times}6$ matrix channel from all six base-station
antennas to its two antennas. The MSs then compress the channel
estimatesw, according to the IEEE802.11ac standard (as described
in Section \ref{feedback}) and sends the result to the BS which
unpacks it. 

\begin{figure}
\centerline{
   \includegraphics[width=0.45\textwidth, height=0.05\textheight ]%
     {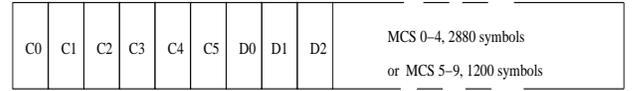}}
\caption{Training block}
\label{frame}
\end{figure}

The BSs then calculates the beamformers as described in Section \ref{beamf},
and sends a second frame using these. 
This frame is sent simultaneously from all the three BSs 20ms after
the first frame. The frame is 3.2ms and consists of six identical
subframes (in order to provide statistics for our evaluation).
The spacing between the subframes is 0.5ms.
Each subframe has two training blocks. Each training block 
is formatted as shown in Figure \ref{frame}. First six symbols
are transmitted sequentially from the six BS transmitter antennas
(C0-C5).
These symbols are not used by the receiver, but will turn
out to be very useful in our analysis of the system.
The following three symbols are training symbols corresponding
to the three simultaneously transmitted streams (D0-D2).
These symbols correspond to 
demodulation reference
symbols in LTE nomenclature and correspond to the VHT-LTF field in 802.11ac.
These symbols are used in the receiver to calculate the weights
of a structured MMSE combiner. Last in the training block are 
the payload symbols (actually the vast majority of symbols). 
In the first training block, these
symbols are encoded using MCS 0 to 4, 
corresponding to rates 1-2.5 bits/subcarrier/symbol.
In the second training block, MCSs with rates of 3-6 bits/subcarrier/symbol
are transmitted. The reason for dividing the subframes into 
two training blocks is to avoid the need for adaptive receiver
algorithms.

In the evaluation of the system, we focus on the sum throughput.
We measure the sum throughput by decoding the received bits and determine
the highest MCS for which there are no
bit errors i.e. when the frame is successfully decoded. By doing
so a practically achievable rate is obtained. Due to the six subframes
in the burst, six throughput measurements are obtained per frame.

In a commercial system, the latency of 20ms is unacceptable. However, a training
repetition time of 23ms could be considered as an option with low overhead.
In such as case, our measurements correspond to the performance at the end of
the update period.

\subsection{Feedback}
\label{feedback}

The standardized feedback scheme of IEEE802.11ac 
is derived in \cite{ROH:07} where it is called ``simple feedback method for slowly time-varying channels''.
In the scheme, the channel matrix is first decomposed
using a singular value decomposition as 
\begin{equation}
{\bf H} = {\bf U} {\bf S}{\bf V}^H.
\end{equation}
In our case, the matrix ${\bf H}$, is the channel
between all six transmitter antennas in the system and the
two antennas at the mobile station.
According to the feedback scheme, the ${\bf V}$
matrix is encoded into two sets of angles, $\phi$ and $\psi$,
using a procedure which is similar to QR decomposition
using Givens rotations.
The number of $\phi$ and $\psi$ angles are 
both $mn-n/2-n^2/2$, where $m$ is the number of BS antennas
and $n$ the number MS antennas. Using $b_{\phi}$ and
$b_{\psi}$ bits respectively to quantize the $\phi$ and $\psi$ angles,
the number of bits needed
in total to encode one ${\bf V}$ matrix is
$((2m-1)n-n^2)(b_{\phi} + b_{\psi})/2$.
The number of bits for the angles $b_{\phi}$ and $b_{\psi}$ (see above),
can have the values $b_{\phi}=5$ and $b_{\psi}=7$ or 
$b_{\phi}=7$ and $b_{\psi}=9$ according to the standard. Herein,
only the latter value has been used.
We refer to \cite{ROH:07} and our previous paper \cite{ZET:14a}
for more details on the procedure of obtaining
the angles (see also our Matlab/Octave implementation at 
\url{http://people.kth.se/~perz/packV/}).
The diagonal ${\bf S}$ matrix is also fed back.
After dividing the diagonal elements of the matrix
with the standard deviation of the noise,
the result corresponds to the square root of the SNR of the corresponding 
stream. These SNRs are quantized and sent to the transmitter
together with the encoding of the ${\bf V}$ matrix.

Since we are using an OFDM modulation scheme, there is actually
one channel matrix per subcarrier. Since adjacent subcarriers
will experience similar channels, it is not necessary to
feed back the channel matrix for every subcarrier.
The IEEE802.11ac standard defined as parameter $N_g$.
When $N_g=4$, the channel is only reported on every fourth
subcarrier. In this paper we will use $N_g=8$ which is actually
not defined in the standard but which turned out to be sufficient
in our environment according to the measurements in \cite{ZET:14a}.

Thus since $N_g=8$ the $\phi$ and $\psi$ values are reported
for every eighth subcarrier which means four ${\bf V}$
matrices in our case. The SNRs are quantized in two steps.
First an average SNR is calculated for all the reported
subcarriers. 
This average is then uniformly quantized with eight bits in the range
from -10dB to 53.75dB. The SNR is then reported with half of the granularity
of the ${\bf V}$ matrix. The SNR on a certain subcarrier is then encoded
as the difference between the average SNR and the SNR of that particular
subcarrier (this is done for each streams individually). 
This value is quantized  between -8dB and 7dB using four bits
i.e. using a one dB granularity. 
%
%
%
The number of feedback bits required (per sub-carrier) is given by 
\begin{multline}
n_{\text{feedback per subcarrier}} = \\
((2m-1)n-n^2)(b_{\phi} + b_{\psi})/(2N_g)+ 4/N_g + 16/N_c,
\end{multline}
where $N_c$ is the total number of subcarriers. In our case, $m=6$, $n=2$, $N_g=8$, $N_c=38$, $b_{\phi}=7$ and $b_{\psi}=9$ this number is 18.921
(719 bits in total).
Assuming a feedback rate of two bits/symbol/subcarrier, 10 OFDM symbols 
are needed to convey this information. This corresponds to a time of 
0.076ms per mobile-station (assuming the feedback is performed in a
TDMA manner) including a 40$\mu$s preamble. 
The IEEE802.11ac standard defines a procedure where several
frames are being exchanged during the feedback procedure.
This procedure is outlined in Section II of \cite{POR:13}.
 The signalling includes null data packet announcement, 
null data packet, beamforming report poll and short inter-frame spacing 
which amounts to an additional 0.366ms.
In total the overhead would thus becomes 0.564ms.
Assuming an update cycle of 23ms, the feedback overhead
is only 2.5\%. With an update rate of 3ms, the overhead
becomes 18.8\%. 
The reason for the large additional overheads 
(null data packet announcement,
null data packet, beamforming report and short inter frame spacing)
is the compatibility with the 
contention based channel access (CS/CSMA) of 802.11 standards.
If the timeslots of these transmissions could be pre-reserved, 
and the header reduced from 40$\mu$s to 3 OFDM symbols, 
the overheads 
would reduce to only 0.7\% and 5.4\%, for an update rate
of 23ms and 3ms, respectively (this would also require that the
pilot symbols of the null data packet where piggy backed
on the last transmission from the BSs).

The paper \cite{POR:13} also investigates further reductions
in the overhead by utilizing the correlation between adjacent
channel estimates, which would be applicable to our scenario as well.

\subsection{Beamformers}
\label{beamf}

In the IA case, the beamformers are first initialized
in the closed form solution given 
in Appendix II of \cite{CAD:08}. 
This solution is then refined using twelve iterations of the 
max-SINR method given in \cite{GOM:08}. This method
aims to optimize the SINR of each of the three users.
This SINR can be expressed as 
\begin{equation}
\text{SINR}_k = \frac{|{\bf u_k}^H {\bf H}_{k,k} {\bf v}_k|^2}
{\sum_{j=1}^{K} |{\bf u_k}^H {\bf H}_{k,j} {\bf v}_j|^2 + \sigma^2},
\label{maxSINR}
\end{equation}
using the notations of \cite{GOM:08}.  
%
%
We employ the iterative max-SINR algorithm also in the CoMP case.
In this case the channel matrices in (\ref{maxSINR}) are of size 
$2{\times}6$,
and only one channel matrix is used per MS.
We initialize the transmit beamformers in the pseudo-inverse
of the eigenbeamformers for the three users.

The beamformers are calculated for the subcarriers where the
channel is fed back, see Section \ref{feedback}, and re-used
on adjacent subcarriers.

\section{RF Impairment Model} 
\label{impairment}

Since we perform measurements of real-world transmission over a wireless
channel, we do not need a propagation model, nor do we need a hardware
impairment model.  
However, the use of RF-impairment models is becoming a popular way of 
bringing realism to simulations and analysis of wireless systems.
We therefore include results,
where we use simulations on the estimated channels from the measurements,
and compare with the actual transmission results from 
the same measurements.
This is done in order to investigate the accuracy of this type of modeling.
In addition, we also use this modeling to extrapolate our results
to a shorter feedback update interval than we have actually implemented,
More on this will follow in Section \ref{ima}.

\begin{figure}
\begin{psfrags}
\psfrag{Ideal}[c][c][1.0]{Ideal signal}
\psfrag{Impair}[c][c][1.0]{Impaired}
\psfrag{signal}[c][c][1.0]{signal}
\psfrag{T}[c][c][1.0]{$\exp(j\theta_m)$}
\psfrag{KTX}[c][l][1.0]{EVM($P$)}
\psfrag{P}[c][c][1.0]{$\sqrt{P}$}
\psfrag{RMS}[c][c][1.0]{RMS}
\psfrag{AWGN}[c][c][1.0]{AWGN}
\centerline{
   \includegraphics[width=0.45\textwidth, height=0.1\textheight ]%
     {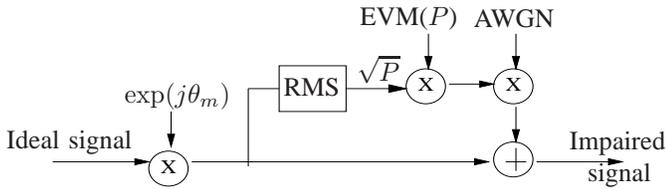}}
\end{psfrags}
\caption{Impairment model for the transmitter}
\label{impair}
\end{figure}

The impairment model used for the transmitter is illustrated by 
Figure \ref{impair} and is similar to the simplified model
in Section 7.2.2 of \cite{SCH:08} with negligible IQ imbalance
(the parameters ${\bf K}_1$ and ${\bf K}_2$ of \cite{SCH:08}
become an identity- and zero-matrix, respectively).

 Some recent work has also used a similar model for finding globally optimal beamformers for multicell MISO networks  \cite{BJO:12b}, and for finding locally optimal precoders for multicell MIMO networks \cite{BRA:14}. However, the model is applied subcarrier-by-subcarrier in  \cite{BJO:12b} and \cite{BRA:14}, for mathematical tractability.


The impaired signal is obtained from the ideal by multiplying
all subcarriers with a common phasor $\exp(j\theta(t))$.
This phase rotation is known as the common phase error (CPE); see
e.g. \cite{SCH:08}, since it affects all subcarriers in the same way.
However, since each of the six transmitters in our system employs
its own PLL-based local oscillator (LO), it can be regarded as independent
between transmitters - which deteriorates beamforming performance.

As shown in Figure \ref{impair} an additive white Gaussian noise (AWGN)
(complex circular symmetric) is added to the received
signal. The power of this noise is governed by the root-mean-square (RMS)
of the input signal and the error-vector-magnitude (EVM) parameter.
The EVM is in general a function of the power of the signal
and is a measure of the modulation accuracy achieved by the transmitter.
The use of a AWGN source is motivated in e.g. \cite{COSTA:02},
and has been used also in \cite{GOR:08}. 
In the receiver we use the same impairment model as in the transmitter, but
without any CPE. Since our receivers use a common LO in its two receiver
branches, the impact of the CPE should be negligible. 
In the sequel, we will sometimes refer to our impairment model as
``the EVM model''.

\begin{figure}
\begin{psfrags}
\psfrag{P}[t][c][1.0]{Output power (dBm)}
\psfrag{E}[c][c][1.0]{EVM {\%}}
\centerline{
   \includegraphics[width=0.36\textwidth, height=0.19\textheight ]%
     {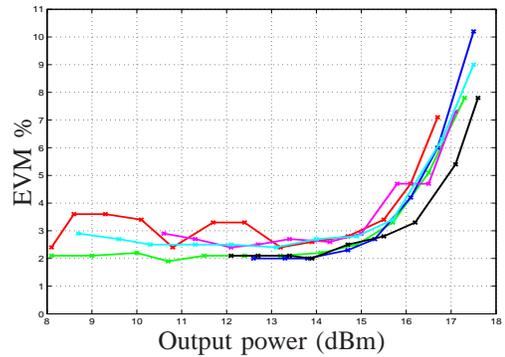}}
\end{psfrags}
\caption{EVM measurements on the transmitters}
\label{EVM_tx}
\end{figure}

In Figure \ref{EVM_tx}
below we have plotted the EVM of our six transmitters as a function
of the output power. The EVM was measured by capturing the output
using a spectrum analyzer. The signal transmitted was a plain OFDM
signal with 16QAM constellation generated according to our parameters.
The received signal was processed in Matlab where channel estimation
was performed first, followed by CPE correction based on the pilot subcarriers.
The EVM is then calculated based on the received constellation points
(see also \url{http://en.wikipedia.org/wiki/Error_vector_magnitude}).
The rise at high power levels is due to the compression in the power amplifiers.
The output power was rarely above 13dBm from any single amplifier 
during the measurements presented in Section \ref{res}.
The standard deviation of the CPE varies from 0.62 degrees to 1.5 degrees.
The distribution of the CPE can be characterized as Laplacian.

\begin{figure}
\begin{psfrags}
\psfrag{P}[t][c][1.0]{Input signal level (dBm)}
\psfrag{E}[c][c][1.0]{EVM {\%}}
\centerline{
   \includegraphics[width=0.36\textwidth, height=0.19\textheight ]%
     {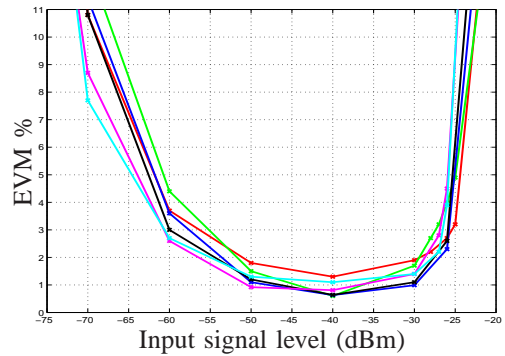}}
\end{psfrags}
\caption{EVM measurements on the receivers}
\label{EVM_rx}
\end{figure}

The EVM of our receivers is shown in Figure \ref{EVM_rx}. These measurements
were performed using a high quality signal generator as input source.
The rise at low
input levels is due to the impact of thermal noise.
The curves shown in Figure \ref{EVM_tx} and  \ref{EVM_rx}; are
first used as look-up tables when performing simulations using the 
impairment model. 
More details of the use of the model is provided
in Section \ref{ima}.

\section{Results}
\label{res}

\subsection{Raw results}
\label{raw}

The measurements were performed as follows. The three MSs were placed
in random locations. The IA scheme is then run first followed by 
CoMP 0.34 seconds
later. After this three reference schemes are run one at a time.
The three reference schemes are ``TDMA MIMO'', ``full-reuse SIMO'' (FR SIMO)
and ``full-reuse MIMO'' (FR-MIMO).
In TDMA MIMO, only one BS-MS pair is active
at a time. But instead two streams are active simultaneously.
In FR-SIMO each base-station transmits simultaneously
using one antenna. In FR-MIMO, all links are
active at the same time using two streams each i.e. six streams
are active in total. None of the reference schemes use beamforming
in the transmitter and therefore no feedback according to
Section \ref{feedback} is needed in these schemes.
All the schemes are run three times in every location.
For every location we also perform measurements with 
one person walking in the measurement area
and with two persons walking in the measurement area.
This gives us the possibility to analyze the influence of
channel variations in a typical office scenario with
increasing mobility.
The three MSs are then moved to new random location and the entire
sequence is repeated.

In total 21 locations were tried. The first 10 were in line-of-sight
(LoS), which means that all three MSs were in the corridor, see Figure \ref{C},
while the last 11 where in NLoS i.e. the three MSs were located in the rooms
adjacent to the corridor.

The performance in LoS and NLoS scenarios is illustrated by the
solid lines in Figure \ref{res3} and \ref{res4}.
The gain of IA and CoMP over the best reference scheme is summarized
in Table \ref{IAGAIN}  (the red italic numbers will be explained
in Section \ref{ima}). The reason for better IA results in NLoS is the 
better isolation of the cells in terms of path-loss, as was
shown in \cite{ZET:14a}.
We note that IA deteriorates with increasing mobility, while it still
able to provided a gain over the reference schemes. The fact that IA
is always better than FR-SIMO also proves that the transmitter beamforming
is effective.
CoMP is more robust to the channel variations than IA.
The reason could be that CoMP will utilize strong channel elements
more than IA. The relative change of weak channel elements is typically
much faster than strong channels (e.g. a channel in a deep fade).

\begin{table} 
\begin{center}
\begin{tabular}{|c|c|c|}\hline
\multicolumn{3}{|c|}{\bf LoS} \\ \hline
 & IA & CoMP  \\ \hline
Stationary & 14\% & 33\%  \\ \hline
One person walking  & 10\%  & 32\%  \\ \hline
Two persons walking  & 3.5\% {\it \color{red} 10\%} & 36\%  \\ \hline
\multicolumn{3}{|c|}{\bf NLoS} \\ \hline
 & IA & CoMP  \\ \hline
Stationary  & 37\% & 59\%  \\ \hline
One person walking  & 32\% & 61\% \\ \hline
Two persons walking  & 26\% {\it \color{red} 33\%} & 50\% 
\\ \hline
\multicolumn{3}{|c|}{\bf LoS+NLoS} \\ \hline
 & IA & CoMP  \\ \hline
Stationary  & 27\% & 47\%  \\ \hline
One person walking  & 23\%  {\it \color{red} 24\%} & 49\% \\ \hline
Two persons walking  & 16\% {\it \color{red} 24\%} & 45\% \\ \hline
\end{tabular}
\end{center}
\caption{Gain of IA and CoMP over best reference with 23ms update rate and 2.5\% overhead deducted. Red italic numbers are predictions of the performance
with 3ms update rate and 5.4\% overhead, see Section \ref{ima}}
\label{IAGAIN}
\end{table}




\subsection{Results with ideal hardware}

In order to investigate the ``would-be'' performance with perfect hardware,
simulations are performed where the transmitted signals are convolved with
the measured channels.
The exact same code (software) is used for the transmitter and receiver 
processing as in the real-time evaluations.  
The channels are estimated from the C0-C5 pilots in the frames (see above).
The same pilot symbols are used throughout the air interface
(orthogonality is achieved through time multiplexing).
This implies that non-linearity
effects should always appear identically at the transmitter
each time the pilot is transmitted.
If this is not the case, the channel estimates will vary
between frames even if the true channel does not - and this
will be erronously appear as channel variations.

\begin{figure}
\begin{psfrags}
\psfrag{T}[c][t][0.8]{Sum throughput (bits/symbol/subcarrier)}
\psfrag{1.2}[c][b][1.0]{}
\psfrag{1.4}[c][b][1.0]{}
\psfrag{1.6}[c][b][1.0]{}
\psfrag{1.8}[c][b][1.0]{}
\psfrag{2.2}[c][b][1.0]{}
\psfrag{2.4}[c][b][1.0]{}
\psfrag{2.6}[c][b][1.0]{}
\psfrag{2.8}[c][b][1.0]{}
\psfrag{1}[t][c][1.0]{No mobility}
\psfrag{2}[t][c][1.0]{One person walking}
\psfrag{3}[tr][c][1.0]{
\begin{minipage}[c]{2cm} \begin{flushright} Two persons 
\end{flushright} \end{minipage}}
\centerline{
   \includegraphics[width=0.45\textwidth, height=0.23\textheight ]%
     {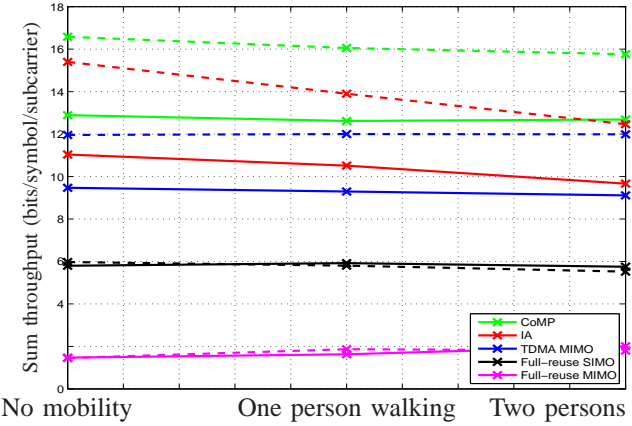}}
\end{psfrags}
\caption{Sum throughput LoS measurements. The dotted lines shows the ``would-be'' 
performance with perfect hardware.}
\label{res3}
\end{figure}

\begin{figure}
\begin{psfrags}
\psfrag{T}[c][t][0.8]{Sum throughput (bits/symbol/subcarrier)}
\psfrag{1.2}[c][b][1.0]{}
\psfrag{1.4}[c][b][1.0]{}
\psfrag{1.6}[c][b][1.0]{}
\psfrag{1.8}[c][b][1.0]{}
\psfrag{2.2}[c][b][1.0]{}
\psfrag{2.4}[c][b][1.0]{}
\psfrag{2.6}[c][b][1.0]{}
\psfrag{2.8}[c][b][1.0]{}
\psfrag{1}[t][c][1.0]{No mobility}
\psfrag{2}[t][c][1.0]{One person walking}
\psfrag{3}[tr][c][1.0]{
\begin{minipage}[c]{2cm} \begin{flushright} Two persons 
\end{flushright} \end{minipage}}
\centerline{
   \includegraphics[width=0.45\textwidth, height=0.23\textheight ]%
     {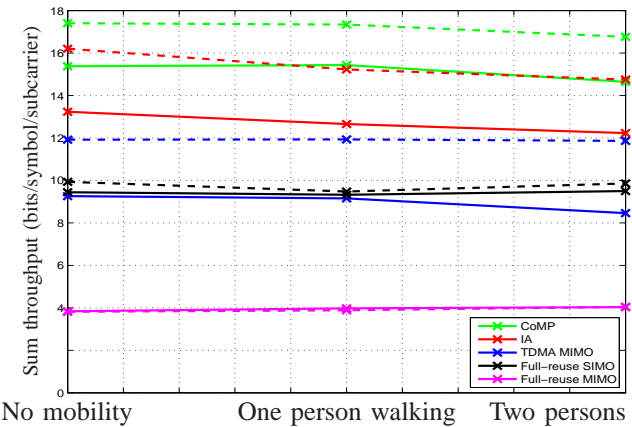}}
\end{psfrags}
\caption{Sum throughput NLoS measurements. The dotted lines shows the ``would-be'' 
performance with perfect hardware.}
\label{res4}
\end{figure}

The result for the LoS and NLoS case is shown by the dashed lines
in Figure \ref{res3}
and \ref{res4}, respectively. 
The LoS results show that sum throughput of IA increases 30-40\% with
perfect hardware, while CoMP and TDMA MIMO improves 25-30\%.
The performance of FR-SIMO and FR-MIMO changes insignificantly.
In the NLoS scenario, the performance improvement is 20\%
for IA, 10\% for CoMP and 30-40\% MIMO. 
We note that IA and MIMO rely on suppression of interference using
a minimal number of antennas, and this may be the reason for their
larger vulnerability to hardware impairments as well as channel variations.

\subsection{Impairment model analysis}
\label{ima}

\begin{figure}
\begin{psfrags}
\psfrag{T}[c][t][0.98]{Sum throughput (bits/symbol/subcarrier)}
\psfrag{1.2}[c][b][1.0]{}
\psfrag{1.4}[c][b][1.0]{}
\psfrag{1.6}[c][b][1.0]{}
\psfrag{1.8}[c][b][1.0]{}
\psfrag{2.2}[c][b][1.0]{}
\psfrag{2.4}[c][b][1.0]{}
\psfrag{2.6}[c][b][1.0]{}
\psfrag{2.8}[c][b][1.0]{}
\psfrag{1}[t][c][1.0]{No mobility}
\psfrag{2}[t][c][1.0]{One person walking}
\psfrag{3}[tr][c][1.0]{
\begin{minipage}[c]{2cm} \begin{flushright} Two persons 
\end{flushright} \end{minipage}}
\centerline{
   \includegraphics[width=0.45\textwidth, height=0.23\textheight ]%
     {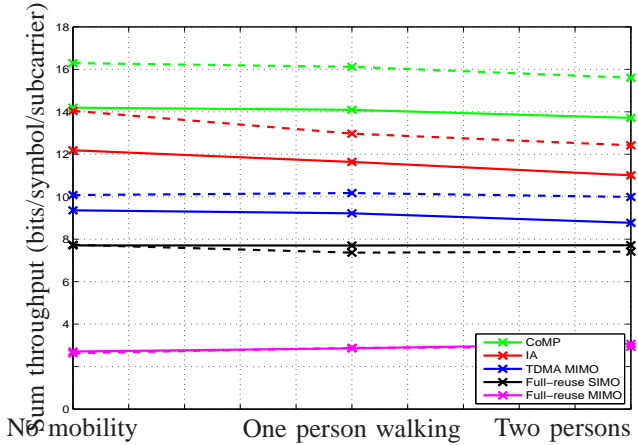}}
\end{psfrags}
\caption{Sum throughput LoS+NLoS measurements. The dotted lines shows the 
results of the EVM-model.}
\label{res5}
\end{figure}

In Section \ref{impairment} we introduced an impairment model.
In previous section we simulated the performance of ideal hardware by
convolving the transmitted signal with the exact channels measured
during the ``real-time'' measurements presented in Section \ref{raw}.
Here we modify this procedure by processing the transmit signals
using the impairment model of Section \ref{impairment}
before convolving with the measured channels. In addition we also
feed the received signal through the impairment model
at the receiver. The CPE is modeled as Laplacian with 1.5degrees
standard deviation based on our worst case measurement.
The RMS value of the signals are calculated per OFDM symbol. 
The EVMs are based look-up tables for each transmitter and
receiver based on the measurements shown in 
Figure \ref{EVM_tx} and Figure \ref{EVM_rx}.
The result (Los and NLoS combined) is shown in Figure \ref{res5}.
The EVM model still over-estimates the performance by
9-21\% in the IA and CoMP cases, 
and 6-14\% in the TDMA-MIMO case.

Using the EVM-model we are able to estimate the performance using 3ms
update rate. We do this by selecting the channel estimates
from the first
subframe for the beamformer calculation (rather than the
channel estimates from the training frame)
\footnote{Note that the update rate is not directly a parameter of the EVM-model
but a consequence of the channel estimates we use for emulating the
propagation channel}.
We then evaluate the performance on all six subframes
using the channels of those subframes.
The results are summarized in Table \ref{movtable},
for different update rates (disregarding overheads).
The results show that with the 3ms update, the performance
can be brought back to that of the stationary channel
using the EVM model. 

Based on this result, we assume that the 
performance of the real system will also return to that
of the stationary case, with the 3ms update rate.
In Section \ref{feedback} we calculated the feedback
overhead according to the IEEE802.11ac standard to
be 2.5\% and 18.8\%, at 23ms and 3ms update intervals,
respectively. The additional overhead of going to a
3ms update interval does not pay off. However, assuming that
some of the overhead signaling of IEEE802.11ac 
could be avoided would bring the overhead down to 5.4\%.
Under this assumption the performance of IA improves 
with the 3ms update rate while CoMP improves only marginally in a
few cases.
The gain of IA with this reconfiguration is 10-33\% as
shown in red in italic Table \ref{IAGAIN}.

\begin{table} 
\begin{center}
\scalebox{0.93}{
\begin{tabular}{|c|c|c|c|}\hline
Scheme & 
\begin{minipage}[c]{1.3cm} 23ms actual \\ stationary \end{minipage} & 
\begin{minipage}[c]{1.3cm} 23ms EVM-model \\ stationary \end{minipage} & 
\begin{minipage}[c]{1.3cm} 3ms EVM-model \\ two persons \\ walking \end{minipage} \\ \hline
\multicolumn{4}{|c|}{LoS} \\ \hline 
IA & 11.0 & 13.0 & 12.8  \\ \hline 
CoMP & 12.9 & 15.3 & 15.8  \\ \hline 
TDMA-MIMO & 9.5 & 10.3 & 10.3  \\ \hline 
FR-SIMO & 5.8 & 5.7 & 5.3 \\ \hline 
FR-MIMO & 1.5 & 1.4 & 1.8   \\ \hline
\multicolumn{4}{|c|}{NLoS} \\ \hline 
IA & 13.2 & 15.0  & 15.1  \\ \hline 
CoMP & 15.4 & 17.2 & 17.3  \\ \hline 
TDMA-MIMO & 9.3 & 9.9 & 9.7  \\ \hline 
FR-SIMO & 9.4 & 9.6 & 9.4 \\ \hline 
FR-MIMO & 3.8 & 3.8 & 4.0   \\ \hline
\multicolumn{4}{|c|}{LoS+NLoS} \\ \hline 
IA & 12.2 & 14.0  & 14.0  \\ \hline 
CoMP & 14.2 & 16.3 & 16.6  \\ \hline 
TDMA-MIMO & 9.4 & 10.1 & 10.0  \\ \hline 
FR-SIMO & 7.7 & 7.7 & 7.4 \\ \hline 
FR-MIMO & 2.7 & 2.6 & 3.0   \\ \hline
\end{tabular}
}
\end{center}
\caption{Sum throughput in bits/symbol/subcarrier using the real-system 
and EVM-model - disregarding feedback overhead}
\label{movtable}
\end{table}

\section{Conclusion}
\label{conc}

We have implemented IA and CoMP on a system consisting of three BSs and
three MSs each having two antennas. The sum throughput of the
system has been studied and compared with reference schemes
such as TDMA MIMO and FR SIMO. Theoretically, at high SNR the gain
of IA over TDMA should be 50\% based on a degrees of freedom analysis.
The closest we get to this number is in the NLoS stationary scenario
where the gain is 27\%. However, this gain drops when there are
people walking in the environment. 
{\em However, there is still a gain even for the time varying channel and with the overheads.}

In the LoS scenario, gains with IA are small. This is due
to the worse path-loss ratios (cell isolation) in this scenario 
as we showed already in \cite{ZET:14a}. 

Coordinated multipoint (CoMP)  provides a gain over the reference schemes
in both LoS and NLoS, 33\% and 59\%, respectively, and seems
to be robust to channel variation induced by walking people.

Our results showed that our hardware reduces the performance
by some 10\%-40\%. It should be kept in mind that the EVMs
measured on our equipment are in the range of what could
be expected on consumer-grade equipment.
An impairment model based on EVM measurements of our testbed hardware
was used and compared with the actual results. The model does
bring us closer to the real system but there is still an overestimation
of the performance. This shows the necessity of incorporating
the impact of RF impairments in the analysis of advanced MIMO
techniques and the importance of experimental performance evaluations.


\bibliographystyle{IEEEtran}
\bibliography{/home/perz/iwssip2012/ref.bib}

\end{document}